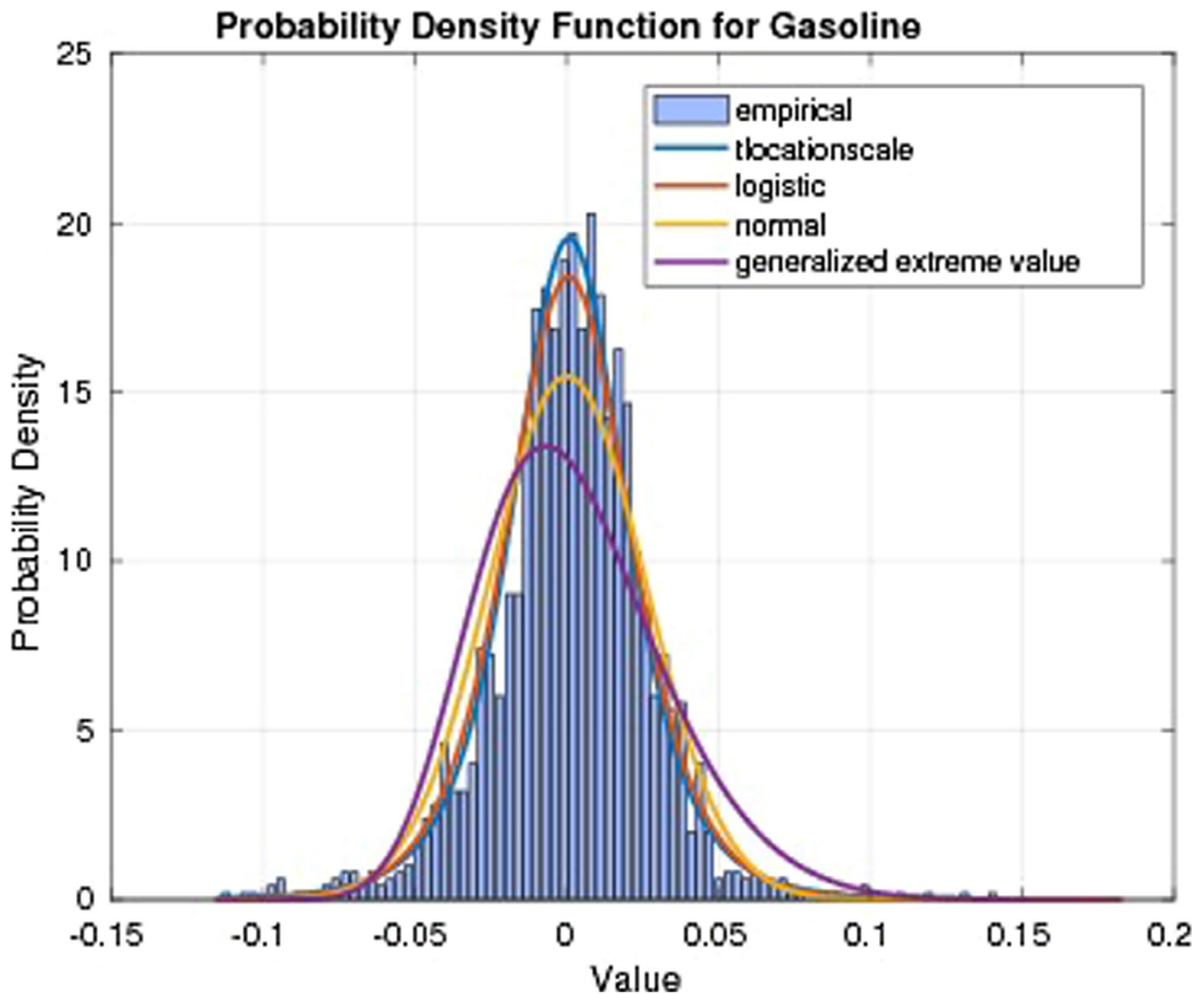

**ECONOMETRICS | RESEARCH ARTICLE**

# Parameter estimation for stable distributions with application to commodity futures log-returns

M. Kateregga, S. Mataramvura and D. Taylor

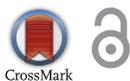

**ECONOMETRICS | RESEARCH ARTICLE**

# Parameter estimation for stable distributions with application to commodity futures log-returns

M. Kateregga[1]*, S. Mataramvura[1] and D. Taylor[1]

**Abstract:** This paper explores the theory behind the rich and robust family of $\alpha$-stable distributions to estimate parameters from financial asset log-returns data. We discuss four-parameter estimation methods including the quantiles, logarithmic moments method, maximum likelihood (ML), and the empirical characteristics function (ECF) method. The contribution of the paper is two-fold: first, we discuss the above parametric approaches and investigate their performance through error analysis. Moreover, we argue that the ECF performs better than the ML over a wide range of shape parameter values, $\alpha$ including values closest to 0 and 2 and that the ECF has a better convergence rate than the ML. Secondly, we compare the *t* location-scale distribution to the general stable distribution and show that the former fails to capture skewness which might exist in the data. This is observed through applying the ECF to commodity futures log-returns data to obtain the skewness parameter.

**Subjects:** Mathematical Finance; Probability; Statistics

## ABOUT THE AUTHOR

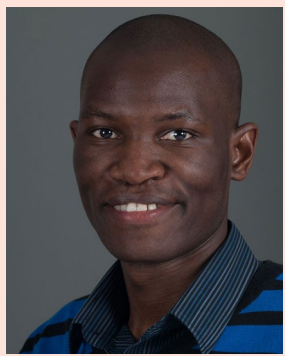

M. Kateregga

Mr M. Kateregga is a finishing PhD student at the University of Cape Town in South Africa. His research is in the field of mathematical finance and his PhD thesis is entitled Stable Distributions with Applications in Finance. The current paper is a chapter in his thesis which is due for submission in August, 2017. Mr Kateregga is also a researcher at the African Collaboration for Quantitative Finance and Risk Research (ACQuFRR) which is the research section of the African Institute of Financial Markets and Risk Management (AIFMRM), which delivers postgraduate education and training in financial markets, risk management and quantitative finance. Mr Kateregga also works with the African Institute for Mathematical Sciences (AIMS) in South Africa as a Research Assistant.

## PUBLIC INTEREST STATEMENT

This paper is entitled parameter estimation for stable distribution with applications to commodity future log-returns. The paper is useful to individuals interested in investing their wealth in financial markets. It provides essential information on how historical asset prices can inform future market movements via parameter estimation. This is crucial to portfolio managers, speculators, and hedgers. It's imperative that the most accurate estimation method is established. Market data distribution deviates from the normal distribution, it exhibits skews, high or low peaks, and fat or skinny tails. The current paper is geared towards establishing the best estimation method among known methods in economic and financial analysis for skewed data.

## 1. Introduction

The motivation for this paper derives from the fact that parameter estimation from historical data is an important analysis to financial market participants. It provides useful information for portfolio managers, speculators, and hedgers. It is therefore, imperative that the most accurate estimation method is established. It is a known fact that in general, market data deviates from the Gaussian distribution, its distribution is either skewed, high or low peaked, and/or with fat or skinny tails. The current paper is geared towards establishing a better parameter estimation method among the commonly known ECF, ML, quantile, and logarithm moments methods used in economic and financial analysis for skewed data assumed to flow stable distributions.

The application of stable distributions in finance is traced way back in the late 50s when Mandelbrot (1959, 1962, 1963) developed a hypothesis that revolutionalized the way economists viewed and interpreted prices in speculative markets such as grains and securities markets. The hypothesis suggested that prices were not Gaussian as it had been previously believed by market participants based on Bachelier (1900). Mandelbrot's hypothesis was therefore, an extension of the widely embraced breakthrough of Bachelier (1900).

In the following years Zolotarev (1964) developed integral representations of stable laws and the results have been used to develop parameter estimation techniques for the stable laws. Fama (1963) reviewed the validity of Mandelbrot's hypothesis and came up with statistical tools suitable for dealing with speculative prices. Dumouchel (1971) employs this class of distributions in statistical inference for long-tailed data. Graphical representation of their densities and the estimation of their parameters via interpolation appear in Holt and Crow (1973) and in Koutrouvelis (1980) using regression. Parameter estimation methods based on quantile methods are presented in Fama and Roll (1971) for symmetric stable distributions but this approach faces a problem of discontinuity of the traditional location parameter in the asymmetrical cases when the exponent parameter passes unity. A remedy and generalization of the quantile approach is later introduced by McCulloch (1986).

A different parameter estimation technique based on fractional lower order moments (FLOM) appears in Ma and Nikias (1995) where the authors develop new methods for estimating parameters in impulsive signal environments. However, their methods only cover symmetric stable distributions. There was a need to extend the method to asymmetric systems. This came through by Kuruoğlu (2001) where a generalized FLOM method is introduced. Generally, FLOM methods pose a challenge of having to estimate the *Sinc* function and this in turn affects the accuracy of the results. As a consequence a better estimation approach referred to as logarithmic moments method (LM) is proposed by Kuruoğlu (2001) to avoid having to compute the *Sinc*.

The third estimation method utilizes the maximum likelihood (ML). It is known that the ML approach is widely favored in economic and financial applications due to its generality and asymptotic efficiency (see for instance, Yu, 2004). However, there are cases where the ML method can be unreliable especially when the likelihood function is not tractable, or its not bounded over the parameter space or does not have a closed form representation. For instance, in this current paper the densities considered do not have closed form expressions. However, since there is a one-to-one correspondence between the density function and its Fourier transform it could be worth exploiting the latter since it always exists and its bounded. This leads us to next estimation method.

The fourth estimation approach is the empirical characteristic function (ECF) method discussed in Yu (2004). Although the likelihood function can be unbounded, its Fourier transform is always bounded and, while the likelihood function might not be tractable or could not be of a closed form, the Fourier transform could have a closed form expression. The Fourier transform of the density function is the characteristic function (CF), hence the name empirical characteristic function (ECF) method. In this paper we aim to show that this approach performs better than all the previously mentioned methods. A useful software package that can be used to estimate stable distributions is provided in Nolan (1997). A more theoretical approach to statistical estimation of the parameters of stable laws is extensively discussed in Zolotarev (1980). Readers interested in how to simulate stable process can refer to two excellent literatures of Weron and Weron (1995) and Zolotarev (1986).

This paper explores the theory behind the rich and robust family of $\alpha$-stable distributions to estimate parameters from financial asset log-returns data. We discuss four-parameter estimation methods including the quantiles, logarithmic moments method, ML, and the empirical characteristics function (ECF) method. The contribution of the paper is two-fold: first, we discuss the above parametric approaches and investigate their performance through error analysis. Moreover, we argue that the ECF performs better than the ML over a wide range of shape parameter values, $\alpha$ including values closest to 0 and 2 and that the ECF has a better convergence rate than the ML. Secondly, we compare the $t$ location-scale distribution to the general stable distribution and show that the former fails to capture skewness which might exist in the data. This is observed through applying the ECF to commodity futures log-returns data to obtain the stable parameters.

The rest of the paper is organized as follows: in Section 2 we define a stable process and its construction from independent and identically distributed random variables based on a generalized central limit theorem and discuss its characterization. In Section 3 we study the density and distribution properties of stable processes through their characteristic functions. Section 4 explains how the four-parameter estimation methods discussed in this paper work and provides an analysis on their accuracy. In Section 5 we study and analyze some commodity data and show that the data deviates from the normal distribution hypothesis. We use the ECF to obtain the four stable parameters from the data and in addition, fit the data to various distributions to determine the closest shape of the data which turns out to be the $t$ location-scale distribution for all our data. This distribution is suited for data that is highly peaked and heavily tailed with outliers. However, we propose stable distribution fitting to check for any existing tails. Section 6 concludes.

## 2. Stable processes

Stable also known as alpha-stable (or equivalently $\alpha$-stable) processes belong to a general class of Lévy distributions. They are limiting distributions with a definitive exponent parameter $\alpha$ that determines their shape.

### 2.1. Definition and construction

*Definition 2.1*  Let $X_1, X_2, \ldots, X_n$ be independent and identically distributed random variables and suppose a random variable $S$ defined by

$$S \Rightarrow \frac{1}{a_n}\left(\sum_{i=1}^{n} X_i - b_n\right),\tag{1}$$

where "$\Rightarrow$" represents weak convergence in distribution, $a_n$ is a positive constant and $b_n$ is real. Then $S$ is a stable process and the constants $a_n$ and $b_n$ need not be finite.

Definition 2.1 allows modeling of a number of natural phenomenon beyond normality using stable distributions. The fact that $a_n$ and $b_n$ do not necessarily have to be finite provides the generalized central limit theorem.

*Definition 2.2*  (Generalized Central Limit Theorem Rachev (2003))  Suppose $X_1, X_2, \ldots$ denotes a sequence of independent and identically distributed random variables and let sequences $a_n \in \mathbb{R}$ and $b_n \in \mathbb{R}^+$. Then we can define a sequence

$$Z_n := \frac{1}{b_n}\left(\sum_{i=1}^{n} X_i - a_n\right)\tag{2}$$

of sums $Z_n$ such that their distribution functions weakly converge to some limiting distribution:

$$P(Z_n < x) \Rightarrow H(x), \quad n \longrightarrow \infty, \tag{3}$$

where $H(x)$ is some limiting distribution.

The traditional central limit theorem assumes finite mean $a := \mathbb{E}[X_i]$ and finite variance $\sigma^2 := \mathrm{Var}[X_i]$ and defines the sequence of sums

$$Z_n := \frac{1}{\sigma\sqrt{n}} \left( \sum_{i=1}^{n} X_i - na \right), \tag{4}$$

such that the distribution functions of $Z_n$ weakly converge to $h^{sG}(x)$:

$$P(x_1 < Z_n < x_2) \Rightarrow \int_{x_1}^{x_2} h^{sG}(x)dx, \quad n \longrightarrow \infty \tag{5}$$

where $h^{sG}(x)$ denotes the standard Gaussian distribution.

$$h^{sG}(x) = \frac{1}{\sqrt{2\pi}} \exp(-x^2/2). \tag{6}$$

Suppose the independent and identically distributed random variables $X_i$ equal to a positive constant $c$ almost surely and the sequences $a_n$ and $b_n$ in (2) are defined by $a_n = (n-1)c$ and $b_n = 1$, then $Z_n$ is also equal to $c$ for all $n > 0$ almost surely. In this case the random variables $X_i$ are mutually independent and as a result, the limiting distribution for the sums $Z_n$ belong to the stable family of distributions by definition. This is one reason why they are regarded as stable.

## 2.2. Parametrization

*Definition 2.3*    A stable distribution is a four-parameter family denoted by $S(\alpha, \beta, \nu, \mu)$:

  (1) $\alpha \in (0, 2]$ is the characteristic exponent responsible for the shape of the distribution.
  (2) $\beta \in [-1, 1]$ is responsible for skewness of the distribution.
  (3) $\nu > 0$ is the scale parameter (it narrows or extends the distribution around $\mu$).
  (4) $\mu \in \mathbb{R}$ is the location parameter (it shifts the distribution to the left or the right).

Suppose a random variable $s$ follows a stable distribution $S(\alpha, \beta, \nu, \mu)$ then the random variable $z = (s - \mu)/\nu$ has the same-shaped distribution as $s$ but with the location parameter $\mu = 0$ and the scale parameter $\nu = 1$. This is another reason why they are referred to as stable, the shape is maintained after any rescaling.

Densities of $\alpha$-stable distributions do not have closed-form representations except for the case of a Gaussian ($\alpha = 2$), Cauchy ($\alpha = 1$, $\beta = 0$) and Inverse Gaussian or Pearson ($\alpha = 0.5$, $\beta = \pm 1$) distributions.

  (1) Gaussian distribution $N(\mu, \sigma^2)$: $S\left(2, 0, \frac{\sigma}{\sqrt{2}}, \mu\right)$.

$$h^G(x) = \frac{1}{\sigma\sqrt{2\pi}} \exp\left(-\frac{(x-\mu)^2}{2\sigma^2}\right); \quad -\infty < x < \infty.$$

  (2) Cauchy distribution: $S(1, 0, \nu, \mu)$.

$$h^C(x) = \frac{1}{\pi} \frac{\nu}{\nu^2 + (1-x)^2}; \quad -\infty < x < \infty.$$

(3) Lévy distribution (Inverse-Gaussian or Pearson): $S(1/2, 1, \nu, \mu)$.

$$h^L(x) = \sqrt{\frac{\nu}{2\pi}}(x - \mu)^{-3/2} \exp\left(-\frac{\nu}{2(x - \mu)}\right); \quad \mu < x < \infty.$$

The densities are generally computed using characteristic functions through transformations such as the Fourier.[1] One can also refer to the work of Zolotarev (1964, 1980, 1986) for straight-forward and easy-to-compute integral representations of stable distribution and density functions. The distribution functions for the different $\alpha$ values have been tabulated in Dumouchel (1971), Fama and Roll (1968) and Holt and Crow (1973).

## 3. Density and distribution properties

### 3.1. Special case

Let $(X_t, t \geq 0)$ denote a Lévy process. The characterization of $X_t$ is deduced from the Lévy-Khintchine formula.

*Definition 3.1* (Lévy-Khintchine & Applebaum, 2004) Let $X = (X_t)_{t \geq 0}$ be a Lévy process. There exist $b \in \mathbb{R}, \sigma \geq 0$ such that the characteristic function of $X$ is given by

$$\Phi(t) := \mathbb{E}[e^{itX}] = \exp\left(itb - \frac{1}{2}\sigma^2 t^2 + \int_{\mathbb{R}-\{0\}} (e^{itx} - 1 - itx1_{|x|<1})m(dx)\right), \tag{7}$$

where $1_{\{\cdot\}}$ is an indicator function and $m$ is a $\sigma$-finite measure satisfying the constraint

$$\int_{\mathbb{R}-\{0\}} \min(1, |x|^2)m(dx) < \infty; \quad \text{alternatively} \quad \int_{\mathbb{R}-\{0\}} \frac{|x|^2}{1 + |x|^2}m(dx) < \infty. \tag{8}$$

*Definition 3.2* (The Lévy-Itô Decomposition Applebaum (2004)) If $X_t$ is a Lévy process, there exist $b \in \mathbb{R}$, a Brownian motion $B_\sigma(t)$ with variance $\sigma \in \mathbb{R}^+$ and an independent Poisson random measure $N$ on $\mathbb{R}^+ \times (\mathbb{R} - \{0\})$ such that, for each $t \geq 0$,

$$X_t = bt + B_\sigma(t) + \int_{|x|<1} x\tilde{N}(t, dx) + \int_{|x|\geq 1} xN(t, dx), \tag{9}$$

where

$$b = \mathbb{E}\left[X_1 - \int_{|x|\geq 1} xN(1, dx)\right]. \tag{10}$$

The compensated compound Poisson random measure is defined by $\tilde{N} = N - t\lambda$ to preserve the martingale property. The Lévy measure $\lambda$ satisfies (8).

A stable distribution can be constructed by setting $\sigma$ to zero in (7) or the second term on the right of (9) to zero and the Lévy measure in (8) to

$$m(dx) = \frac{C}{|x|^{1+\alpha}}\,dx; \quad C > 0,$$  (11)

This gives a pure jump Lévy process which is a simple example of a stable family of distributions. We discuss a general case in the following.

### 3.2. General case

In the following, $(S_t)_{t\geq0}$ will represent a stable process. Its characteristic function $\Phi$ is obtained using the definition of domain of attraction of stable random variables and the Lévy-Khinchine representation formula in Definition 3.1 (see Applebaum, 2004):

$$\Phi(\theta) = \begin{cases} \exp\left(-\nu^\alpha|\theta|^\alpha\left(1 - i\beta\,\text{sign}(\theta)\tan\left(\frac{\pi\alpha}{2}\right)\right) + i\mu\theta\right); & \text{for } \alpha \neq 1. \\ \exp\left(-\nu|\theta|\left(1 + i\beta\,\text{sign}(\theta)\frac{2}{\pi}\log|\theta|\right) + i\mu\theta\right); & \text{for } \alpha = 1. \end{cases}$$  (12)

Alternative forms of parametrization are discussed in McCulloch (1986) for easier numerical analysis. More discussion on this to follow in Section 3.4.

The density of $S_t$ is computed from (12) using the Fourier transform:

$$h_{S_t}(s) = \frac{1}{2\pi}\int_{-\infty}^{\infty} e^{-ist}\Phi(t)dt.$$  (13)

Figure 1 shows density graphs for different exponent parameter values. The density is defined over the whole real line and for application purposes in finance log-returns data is usually used instead of raw asset prices to fit this family of distributions.

**Figure 1. $\alpha$-stable densities for $\alpha \in (0, 2]$.**

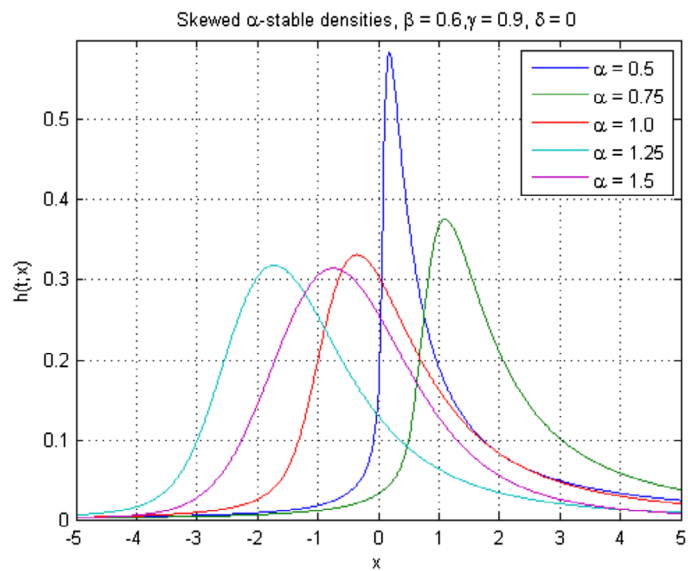

The drawback in approximating (13) is that elementary techniques such as expressing the integral in terms of simple functions or using infinite polynomial expressions of the density function are not sufficient for meaningful numerical analysis. Some authors propose a standard parameterized integral expression of the density given by (see Ament & O'Neal, 2016)

$$h_{S_t}(\alpha, \beta, \nu, \mu) = \frac{1}{\sigma\pi} \int_0^\infty e^{-t^\alpha} \cdot \cos\left(t \cdot \left(\frac{s-\mu}{\sigma}\right) - \beta t^\alpha \tan\left(\frac{\pi\alpha}{2}\right)\right) dt. \tag{14}$$

However, this representation consists of an oscillating integrand which in turn leads to another alternative approach presented in Zolotarev (1986) where the density of $S_t$ is given by

$$h_{S_t}(\alpha, \beta, \nu, \mu) = \begin{cases} \frac{\alpha\left|\frac{s-\mu}{\sigma}\right|^{\frac{1}{\alpha-1}}}{2\sigma|\alpha-1|} \int_{-\theta}^{1} U_\alpha(\varphi, \theta) \exp\left(-\left|\frac{s-\mu}{\sigma}\right|^{\frac{\alpha}{\alpha-1}} U_\alpha(\varphi, \theta)\right) d\varphi; & \text{if } s \neq \mu \\ \frac{1}{\pi\sigma} \cdot \Gamma\left(1 + \frac{1}{\alpha}\right) \cdot \cos\left(\frac{1}{\alpha} \arctan\left(\beta \cdot \tan\left(\frac{\pi\alpha}{2}\right)\right)\right); & \text{if } s = \mu \end{cases} \tag{15}$$

$$U_\alpha(\varphi, \vartheta) = \left(\frac{\sin\left(\frac{\pi}{2}\alpha(\varphi+\vartheta)\right)}{\cos\left(\frac{\pi\varphi}{2}\right)}\right)^{\frac{\alpha}{1-\alpha}} \cdot \left(\frac{\cos\left(\frac{\pi}{2}(\alpha-1)\varphi+\alpha\vartheta\right)}{\cos\left(\frac{\pi\varphi}{2}\right)}\right), \tag{16}$$

where $\theta = \arctan\left(\beta\tan\frac{\pi\alpha}{2}\right)\frac{2}{\alpha\pi}\text{sign}(s-\mu)$.

### 3.3. Some properties of stable distribution functions

Firstly, recall that for any two admissible sets of parameters of stable distributions we can find two unique numbers $a > 0$ and $b$ such that

$$S(\alpha, \beta, \nu, \mu) \stackrel{d}{=} aS(\alpha, \beta, \nu', \mu'), \tag{17}$$

where

$$a = \frac{\nu}{\nu'}, \quad b = \begin{cases} \mu - \mu'\frac{\nu}{\nu'}, & \alpha \neq 1 \\ \mu - \mu'\frac{\nu}{\nu'} + \nu\beta\frac{2}{\pi}\log\frac{\nu}{\nu'}, & \alpha = 1. \end{cases} \tag{18}$$

The intuition is that a general stable distribution can be expressed in terms of a standard stable distribution. That is, we can write $S(\alpha, \beta, \nu, \mu) \stackrel{d}{=} aS(\alpha, \beta, 1, 0) + b$ where

$$a = \nu, \quad b = \begin{cases} \mu; & \alpha \neq 1 \\ \mu + \nu\beta\frac{2}{\pi}\log\nu; & \alpha = 1. \end{cases} \tag{19}$$

Secondly, suppose $h$, $H$ and $\Phi$ denote the respective probability, cumulative density and characteristic functions of a stable random variable, $S$, where

$$h(s, \alpha, \beta) = \frac{1}{2\pi} \int_{-\infty}^{\infty} (\cos st - i \sin st)\Phi(t, \alpha, \beta)dt,$$

then it is readily seen that the following properties hold:

(1) $h(-s, \alpha, \beta) = h(s, \alpha, -\beta)$.
(2) $H(-s, \alpha, \beta) = 1 - H(s, \alpha, -\beta)$.
(3) $\Phi(-s, \alpha, \beta) = \Phi(s, \alpha, -\beta)$.

The above three relations can be verified by trigonometric properties.

### 3.4. Simulating $\alpha$-stable random variables

The two excellent references for simulating stable processes are Zolotarev (1986) and Chambers, Mallows, and Stuck (1976).

*Definition 3.3*   Suppose $S_t$ is a stable process with parameters $(\alpha, \beta_2, \nu_2, \mu)$, the characteristic function is given by

$$\ln \Phi(t) = \begin{cases} i\mu t - \nu_2^\alpha |t|^\alpha \exp(-i\beta_2 \, sign(t)) \frac{\pi}{2} K(\alpha)), & \alpha \neq 1; \\ i\mu t - \nu_2 |t| \left( \frac{\pi}{2} + i\beta_2 \, sign(t) \right) \ln |t| \right); & \alpha = 1; \end{cases} \tag{20}$$

where

$$K(\alpha) = \alpha - 1 + sign(1 - \alpha) = \begin{cases} \alpha; & \alpha \neq 1 \\ \alpha - 2; & \alpha = 1 \end{cases} \tag{21}$$

$$(\beta_2, \nu_2) = \begin{cases} \frac{2}{\pi K(\alpha)} \tan^{-1}\left( \beta \tan \frac{\pi \alpha}{2} \right), \nu \left( 1 + \beta^2 \tan^2 \frac{\pi \alpha}{2} \right)^{\frac{1}{2\alpha}}; & \alpha \neq 1 \\ \beta, \frac{2}{\pi} \nu; & \alpha = 1 \end{cases} \tag{22}$$

LEMMA 3.4   *Let $\gamma \in \left[ -\frac{\pi}{2}, \frac{\pi}{2} \right]$ be a uniformly distributed random variable and let $W$ be an independent exponential random variable with mean 1. Then*

$$S = \begin{cases} \frac{\sin \alpha \left( \gamma + \frac{\pi}{2} \beta_2 \frac{K(\alpha)}{\alpha} \right)}{(\cos \gamma)^{\frac{1}{\alpha}}} \left( \frac{\cos \left( \gamma - \alpha \left( \gamma + \frac{\pi}{2} \beta_2 \frac{K(\alpha)}{\alpha} \right) \right)}{W} \right)^{\frac{1-\alpha}{\alpha}}; & \alpha \neq 1 \\ \left( \frac{\pi}{2} + \beta_2 \gamma \right) \tan \gamma - \beta_2 \log \left( \frac{W \cos \gamma}{\frac{\pi}{2} + \beta_2 \gamma} \right); & \alpha = 1 \end{cases} \tag{23}$$

*is a standard $\alpha$-stable process with parameters $(\alpha, \beta_2, 1, 0)$.*

*Proof*   See Zolotarev (1986). □

A stable random variable can be easily generated using Lemma 3.4. Programming languages such as R or MATLAB can be utilized to generate a uniformly distributed random variable $U$ on the interval $\left( -\frac{\pi}{2}, \frac{\pi}{2} \right)$ and an independent exponential random variable $E$ with mean 1[2]. Then the stable random variable would be generated by computing

$$S = \begin{cases} A_{\alpha, \beta} \frac{\sin(\alpha(U + B_{\alpha, \beta}))}{(\cos U)^{\frac{1}{\alpha}}} \left( \frac{\cos(U - \alpha(U + B_{\alpha, \beta}))}{E} \right)^{\frac{1-\alpha}{\alpha}}; & \alpha \neq 1 \\ \frac{2}{\pi} \left( \left( \frac{\pi}{2} + \beta U \right) \tan U - \beta \log \left( \frac{\frac{\pi}{2} E \cos U}{\frac{\pi}{2} + \beta U} \right) \right); & \alpha = 1 \end{cases} \tag{24}$$

where $A_{\alpha, \beta} = \left( 1 + \beta^2 \tan^2 \frac{\pi \alpha}{2} \right)^{\frac{1}{2\alpha}}$ and $B_{\alpha, \beta} = \frac{\tan^{-1}\left( \beta \tan \frac{\pi \alpha}{2} \right)}{\alpha}$.

### 3.5. Moments of stable processes

Statistical moments $\mathbb{E}[|\cdot|^k]$ of stable distributions are finite only when $k \leq \alpha$. Moreover, for $\alpha < 2$ the variance is infinite, for $\alpha \in (0, 1]$ the mean does not exist and the mean is zero when $\alpha \in (1, 2)$. This is not always the case for symmetric stable distributions where $\beta = 0$.

#### 3.5.1. Fractional lower order moments

The FLOM is an alternative for computing moments of $\alpha$-stable random variables especially in situations where the mean and/or variance are infinite. FLOM representation formulas are discussed in Ma and Nikias ([1995](#)) for symmetric stable random data and its generalization to asymmetric stable random data in Kuruoğlu ([2001](#)). In the latter, if $S_i \sim S(\alpha, \beta, \nu, \gamma)$ and $\alpha \neq 1$, then

$$\mathrm{E}[S^{<p>}] = \frac{\Gamma\left(1 - \frac{p}{\alpha}\right)}{\Gamma(1-p)}\left|\frac{\gamma}{\cos\theta}\right|^{\frac{p}{\gamma}} \frac{\sin\left(\frac{p\theta}{\alpha}\right)}{\sin\left(\frac{p\pi}{2}\right)}, \quad \text{for} \quad p \in (-2, -1) \cup (-1, \alpha).$$

$$\mathrm{E}[|S|^p] = \frac{\Gamma\left(1 - \frac{p}{\alpha}\right)}{\Gamma(1-p)}\left|\frac{\gamma}{\cos\theta}\right|^{\frac{p}{\gamma}} \frac{\cos\left(\frac{p\theta}{\alpha}\right)}{\cos\left(\frac{p\pi}{2}\right)}, \quad \text{for} \quad p \in (-1, \alpha).$$

where $\theta = \arctan\left(\beta \tan\frac{\alpha\pi}{2}\right)$ and $\Gamma$ denotes the Gamma function. From the above representations, moments with negative values of $p$ are attainable. This results into the logarithmic moments approach that provides an easier way of estimating stable distribtuion parameters compared to the FLOM.

### 3.5.2. Logarithmic moments

This approach is as a result of the challenges encountered when using the FLOM method which requires computing Gamma functions, the inversion of the *sinc* function and it only works for some $p$. The current method suggests computing derivatives with respect to the moment order $p$ resulting in moments of the logarithms of the stable process. We illustrate in the following.

LEMMA 3.5   *Let S denote a symmetric stable random variable and let $p \in \mathbb{R}$. Then*

$$M_n := \mathrm{E}[(\log|S|)^n] = \lim_{p \to 0} \frac{d^n}{dp^n}\mathrm{E}[|S|^p], \quad n = 1, 2, \ldots. \tag{25}$$

The moments follow readily for $n = 1, 2, \ldots$ i.e.

$$M_1 = \mathrm{E}[\log|S|] = \varphi_0\left(1 - \frac{1}{\alpha}\right) + \frac{1}{\alpha}\log\left|\frac{\nu}{\cos\theta}\right|. \tag{26}$$

$$M_2 = \mathrm{E}[(\log|S| - \mathrm{E}[\log|S|])^2] = \varphi_1\left(\frac{1}{2} + \frac{1}{\alpha^2}\right) - \frac{\theta^2}{\alpha^2}. \tag{27}$$

$$M_3 = \mathrm{E}[(\log|S| - \mathrm{E}[\log|S|])^3] = \varphi_3\left(1 - \frac{1}{\alpha^3}\right). \tag{28}$$

where $\theta = \arctan(\beta \tan \alpha\pi/2)$ and terms $\varphi_k$ are given by $\varphi_0 = -0.57721566, \varphi_1 = \pi^2/6, \varphi = 1.2020569$ derived from the polygamma function

$$\varphi_{k-1} = \frac{d^k}{dx^k}\log\Gamma(x)|_{x=1}. \tag{29}$$

*Proof*   3.6 The proof is provided in Kuruoğlu ([2001](#)).   □

## 4. Parameter estimation of stable processes

The four common methods for estimating parameters of stable processes include: quantiles method (see Fama & Roll, [1971](#); McCulloch, [1986](#), [1996](#)), the logarithmic moments method (see Kuruoğlu, [2001](#)), the empirical characteristics method (see Yang, [2012](#)), and the ML method (see Nolan, [2001](#)). We investigate their accuracy in the following.

### 4.1. The quantiles method

The quantile method was pioneered by Fama and Roll ([1971](#)) but was much more appreciated through McCulloch ([1986](#)) after its extension to include asymmetric distributions and for cases where $\alpha \in [0.6, 2]$ unlike the former approach that restricts it to $\alpha \geq 1$.

Suppose $\hat{s}$ is a given data sample then the estimates for $\alpha$ and $\beta$ are given by $\hat{\alpha} = \Theta_1(\hat{\vartheta}_\alpha, \hat{\vartheta}_\beta)$ and $\hat{\beta} = \Theta_2(\hat{\vartheta}_\alpha, \hat{\vartheta}_\beta)$ where

$$\hat{\vartheta}_\alpha = \frac{\hat{s}_{0.95} - \hat{s}_{0.05}}{\hat{s}_{0.75} - \hat{s}_{0.25}}, \qquad \hat{\vartheta}_\beta = \frac{\hat{s}_{0.95} + \hat{s}_{0.05} - 2\hat{s}_{0.05}}{\hat{s}_{0.95} - \hat{s}_{0.05}}. \tag{30}$$

The notation $\hat{s}_q$ represents the $q$th quantile of sample $\hat{s}$ and, $\hat{\alpha}$ and $\hat{\beta}$ are obtained by functions $\Theta_1(\hat{\vartheta}_\alpha, \hat{\vartheta}_\beta)$ and $\Theta_2(\hat{\vartheta}_\alpha, \hat{\vartheta}_\beta)$ given in Tables III and IV in McCulloch ([1986](#)) through linear interpolation. Consequently, the scale parameter is given by

$$\hat{v} = \frac{\hat{s}_{0.75} - \hat{s}_{0.25}}{\Theta_3(\hat{\alpha}, \hat{\beta})}, \tag{31}$$

where $\Theta_3(\hat{\alpha}, \hat{\beta})$ is given by Table V in McCulloch ([1986](#)). The consistent estimator $v$ is then obtained through interpolation.

Finally the location parameter $\mu$ is estimated through a new parameter defined by

$$\zeta = \begin{cases} \mu + \beta\gamma \tan\frac{\pi\alpha}{2}; & \alpha \neq 1 \\ \mu; & \alpha = 1. \end{cases} \tag{32}$$

Moreover, $\zeta$ is estimated by

$$\hat{\zeta} = \hat{s}_{0.5} + \hat{v}\Theta_5(\hat{\alpha}, \hat{\beta}), \tag{33}$$

where $\Theta_5(\hat{\alpha}, \hat{\beta})$ is obtained from Table VII (McCulloch, [1986](#)) by linear interpolation. The location parameter is estimated consistently by

$$\hat{\mu} = \hat{\zeta} + \hat{\beta}\hat{v} \tan\frac{\pi\hat{\alpha}}{2}. \tag{34}$$

### 4.2. Empirical characteristic function method

Suppose a set of observable data $\{s_1, s_2, \ldots, s_N\}$ follows a stable distribution. Then we can approximate the characteristic function of this data by applying a basic Monte Carlo approach based on the law of large numbers i.e.

$$\Phi(u) = \mathrm{E}[e^{ius_j}] \approx \hat{\Phi}(u) = \frac{1}{N}\sum_{j=1}^{N} e^{ius_j}. \tag{35}$$

We can express the characteristic function (12) in terms of the cosine and sine function from basic trigonometric principles, i.e.

$$\Phi(u) = e^{-|vu|^\alpha}(\cos\eta + i\sin\eta), \tag{36}$$

where

$$\eta = vu - |vu|^\alpha \beta \, \mathrm{sign}(u)\omega(u, \alpha)$$

$$\omega(u, \alpha) = \begin{cases} \tan\frac{\pi\alpha}{2}, & \alpha \neq 1 \\ \frac{2\log|u|}{\pi}, & \alpha = 1 \end{cases}$$

As a result, we observe that

$$|\Phi(u)| = e^{-|vu|^{\alpha}}. \tag{37}$$

The estimated characteristic function relates to the model parameters by

$$\log|\hat{\Phi}(u_k)| = v^{\alpha}|u_k|^{\alpha}; \text{ for } k = 1, 2, \ u_k > 0, \ \alpha \neq 1. \tag{38}$$

Solving this system leads to the estimation representation formulas for the stability and variance parameters:

$$\hat{\alpha} = \frac{\log \frac{\log|\hat{\Phi}(u_1)|}{\log|\hat{\Phi}(u_2)|}}{\log\left|\frac{u_1}{u_2}\right|}.$$

$$\log \hat{v} = \frac{\log|u_1|\log(-\log|\hat{\Phi}(u_2)|) - \log|u_2|\log(-\log|\hat{\Phi}(u_1)|)}{\log\left|\frac{u_1}{u_2}\right|}.$$

The real and imaginary parts of the characteristic function (36) provide estimates for $\hat{\beta}$ and $\hat{\mu}$:

$$\arctan\frac{Im(\Phi(u))}{Re(\Phi(u))} = \mu u - |vu|^{\alpha}\beta \operatorname{sign}(u)\omega(u, \alpha). \tag{39}$$

Suppose $\Upsilon(u) := \arctan(Im\Phi(u)/Re\Phi(u))$ and choose another set of positive numbers $u_k$, $k = 3, 4$ together with $\hat{\alpha}$ and $\hat{v}$ then the estimates of the location and skewness parameters are given respectively by

$$\hat{\mu} = \frac{u_4^{\hat{\alpha}}\Upsilon(u_3) - u_3^{\hat{\alpha}}\Upsilon(u_4)}{u_3 u_4^{\hat{\alpha}} - u_3^{\hat{\alpha}} u_4}. \tag{40}$$

$$\hat{\beta} = \frac{u_4\Upsilon(u_3) - u_3\Upsilon(u_4)}{\hat{v}^{\hat{\alpha}}\tan\frac{\pi\hat{\alpha}}{2}(u_4 u_3^{\hat{\alpha}} - u_3 u_4^{\hat{\alpha}})}. \tag{41}$$

Notice, it can be deduced from Equation (36) that

$$\log(-\log(|\Phi(u)|^2)) = \log(2v^{\alpha}) + \alpha\log(u).$$

This provides an alternative way to envision the regression estimation method:

$$y_k = m + \alpha x_k + \varepsilon_k; \quad k = 1, 2, \ldots, M;$$

where $y_k = \log(-\log|\hat{\Phi}(u_k)|^2)$, $m = \log(2v^{\alpha})$, $x_k = \log(u_k)$ and $\varepsilon_k$ is an error term. The stability parameter $\alpha$ and the scale parameter $v$ can be estimated by selecting $u_k = \frac{\pi k}{25}$, $k = 1, 2, \ldots, M$; of real data (see Koutrouvelis, 1980, Table I). The estimates $\hat{\alpha}$ and $\hat{v}$ are then used to estimate $\beta$ and $\mu$ using the following relation

$$z_l = \eta_l + \varsigma_l, \quad l = 1, 2, \ldots, Q.$$

where $z_l = \Upsilon_n(u_l) + \pi k_n(u_l)$, $\eta_l = \hat{v}_l u - |\hat{v}_l u|^{\hat{\alpha}}\beta \operatorname{sign}(u)\omega(u, \hat{\alpha})$ and $\varsigma_l$ is some random error. The proposed real data set for $Q$ (see Koutrouvelis, 1980, Table II) is $u_l = \frac{\pi l}{50}$, $l = 1, 2, \ldots, Q$.

### 4.3. Logarithmic moments method
This approach follows the theory discussed in Section 3.5.2. The key innovation with this method is that there is no need of computing Gamma functions and the *sinc* function as in the FLOM. Secondly, techniques of parameter estimation for symmetric stable random variables (i.e. $\beta = 0$) can be applied to skewed stable random variables (i.e. $\beta \neq 0$) and, techniques of parameter estimation for centered stable random variables (i.e. $\mu = 0$) to non-centered ones (i.e. $\mu \neq 0$) through centro-symmetrization. However, this comes at a cost of losing almost half of the sample data. Therefore to obtain better estimates one has to use large sample data sets.

### 4.3.1. Centro-symmetrization of stable random data sets

Let $S_k$ be a sequence of $n$ independent stable random variables distributed according to

$$S_k \sim S(\alpha, \beta, \nu, \mu).$$

Then the distribution of a weighted sum of the above sequence with weights $a_k$ can be estimated using their characteristic function:

$$Z = \sum_{k=1}^{n} a_k S_k \sim S\left(\alpha, \frac{\sum_{k=1}^{n} a_k^{<\alpha>}}{\sum_{k=1}^{n} |a_k|^\alpha} \beta, \sum_{k=1}^{n} |a_k|^\alpha \nu, \sum_{k=1}^{n} a_k \mu\right), \tag{42}$$

where the $p$th power of a number $x$ is defined by

$$x^{<p>} = \text{sign}(x)|x|^p.$$

As a result, it is easy to obtain sequences of independent stable random variables with zero $\mu$, zero $\beta$ as well as both zero $\mu$ and zero $\beta$ for $\alpha \neq 1$. This yields the centred, deskewed, and symmetrized sequences:

$$S_k^C = S_{3k} + S_{3k-1} - 2S_{3k-2} \sim S\left(\alpha, \left[\frac{2 - 2^\alpha}{2 + 2^\alpha}\right]\beta, [2 + 2^\alpha]\nu, 0\right), \tag{43}$$

$$S_k^D = S_{3k} + S_{3k-1} - 2^{1/\alpha}S_{3k-2} \sim S(\alpha, 0, 4\nu, [2 - 2^{1/\alpha}]\mu), \tag{44}$$

$$S_k^S = S_{2k} - S_{2k-1} \sim S(\alpha, 0, 2\nu, 0). \tag{45}$$

### 4.3.2. Parameter estimation

Suppose $S_k$ is a data set assumed to be drawned from $S(\alpha, \beta, \nu, \mu)$. Then the exponent parameter $\alpha$ is estimated by setting $\theta = 0$ in (27), and the log moment $M_2$ is estimated from the obverted data (45). That is,

$$\hat{\alpha} = \left(\frac{M_2}{\varphi_1} - \frac{1}{2}\right)^{-1/2}. \tag{46}$$

The estimated $\hat{\alpha}$ is used to estimate $\theta$ using (26) where $M_1$ is estimated from the obverted data (44). That is,

$$|\hat{\theta}| = \left(\left(\frac{\varphi_1}{2} - M_2\right)\hat{\alpha}^2 + \varphi_1\right)^{1/2}. \tag{47}$$

From the definition of $\theta$, $|\beta_0|$ can be estimated by

$$\hat{\beta}_0 = \frac{\tan\hat{\theta}}{\tan\frac{\hat{\alpha}\pi}{2}}. \tag{48}$$

Centering (see (43)) requires $|\hat{\beta}_0|$ to be multiplied by $(2 + 2^\alpha)/(2 - 2^\alpha)$ to obtain $|\hat{\beta}|$ of the original data where the sign of $\beta$ is determined by

cogent •• economics & finance

$$K = \text{sign}(|S_{max} - S_{md}| - |S_{min} - S_{md}|), \quad \text{such that} \quad \hat{\beta} = K|\hat{\beta}|.$$

where $S_{max}, S_{md}, S_{min}$ is the maximum, median and minimum of the original data.

Next we estimate the scale parameter $\hat{v}_0$ using (26) where $M_1$ is estimated from the obverted data (43). That is

$$\hat{v}_0 = |\cos\hat{\theta}| \exp((M_1 - \varphi_0)\hat{\alpha} + \varphi_0). \tag{49}$$

Again centering (see (43)) gives the parameter estimate $\hat{v}$ of the original data by $\hat{v} = \hat{v}_0(2 - 2^{1/\alpha})^{-1}$.

Finally, the location parameter $\mu$ is estimated by

$$\hat{\mu} = \hat{\mu}_0(2 - 2^{1/\alpha})^{-1}. \tag{50}$$

where $\mu_0$ is the median or mean of the obverted data ().

### 4.4. Maximum likelihood method

The ML method is the most favored parameter estimation method in economic and financial applications. The method relies on the density function which in the case of stable distributions poses a closed form representation problem. In this case we propose a numerical estimation of the density function. For a vector $s = (s_1, s_2, \ldots, s_n)$ of independent identically distributed random variables assumed to follow a stable distribtion, the ML estimate of the parameter vector $\Theta = (\alpha, \beta, v, \mu)$ is obtained by maximizing the log-likelihood function given by

$$L_{\Theta}(s) = \sum_{i=1}^{n} \ln \tilde{h}(s_i; \Theta), \tag{51}$$

where $\tilde{h}(s; \Theta)$ denotes a numerically estimated stable probability density function. It is shown for instance in Mittnik, Rachev, Doganoglu, and Chenyao (1999) that the best algorithms to compute the ML is by using Fast Fourier Transforms (FFT) or by direct integration method as in Nolan (2001). The ML algorithms require carefully chosen initial input parameters which in our case can be obtained for example, through the quantiles method described above. The FFT is faster for large data sets and the direct integral approach is suitable for smaller data sets since it can be evaluated at any arbitrary point.

In the following section, we analyze commodities and apply the empirical characteristic functions method to estimate the stable distribution parameters.

It is important to mention the restrictions on the parameters under which the different estimation methods operate.

### 4.5. Error analysis

In this section we simulate datasets from the stable family of distributions based on the theory in Chambers et al. (1976) and Weron and Weron (1995). Then use the above four methods to retrieve the stable parameters from the simulated data. Our focus is on the $\alpha$ and $\beta$ but the arguments extend to the other two parameters.

| Table 1. Estimation methods and their parameter restrictions | |
|---|---|
| **Estimation method** | **Parameter restrictions** |
| Quantile | $\alpha \geq 0.1$ |
| Logarithm moments | $\beta = \mu = 0$ |
| Maximum likelihood | $\alpha \geq 0.4$ |
| Empirical characteristic function | $\alpha \geq 0.1$ |

First, it is important to mention that all the four methods perform poorly close to the boundaries i.e. $\alpha \to 0$, $\alpha \to 2$ and $\beta \to \pm 1$. Moreover, literature shows that the methods operate efficiently under the parameter restrictions in Table 1.

In addition, the MLE seems the most preferred and used estimation method. However, we observe in our analysis that this method fails for particular parameter ranges and it is not robust. For instance in estimating $0.1 < \alpha < 1.0$ with respect to $\beta$, the MLE fails to converge and returns huge unrealistic errors. This is why we do not include it in Figure 2(a). Similarly, for $\beta = 0.4$ estimation with respect to $\alpha$, the logarithm moments method returns either negative or very large $\beta$ values which is expected according to the constraints in Table 1. We omit its graph in Figure 2(b). Meanwhile, we notice that in both cases, the quantile and ECF methods work well with the latter providing relatively the best estimates.

**Figure 2. Method comparison for $\alpha = 0.4$ and $\beta = 0.4$ estimation.**

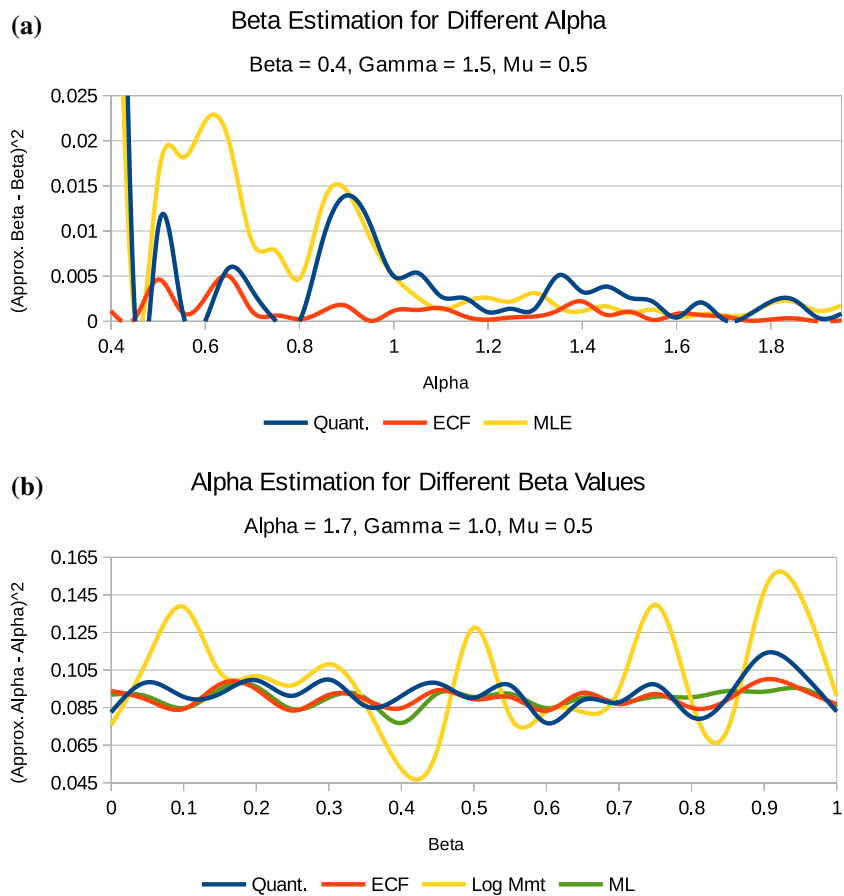

The graphs in Figure 3 show the error associated with estimating $1.0 < \alpha < 2.0$ for different $\beta$ values. Note that all the four methods work well and we still notice the ECF being relatively the most accurate and robust method. Recall that for $\alpha \to 1$ and $\alpha \to 2$ the estimation methods perform poorly. An example is Figure 3(a) (for $\alpha = 1.4$) which was the closest for which the ML would converge but for higher $\alpha > 1.4$ values but far less than 2.0 (see for instance, Figure 3(a) for $\alpha = 1.7$) the methods performed relatively better except for the logarithm moments methods.

**Figure 3. Method comparison for $\alpha = 1.4$ and $\alpha = 1.7$ estimation.**

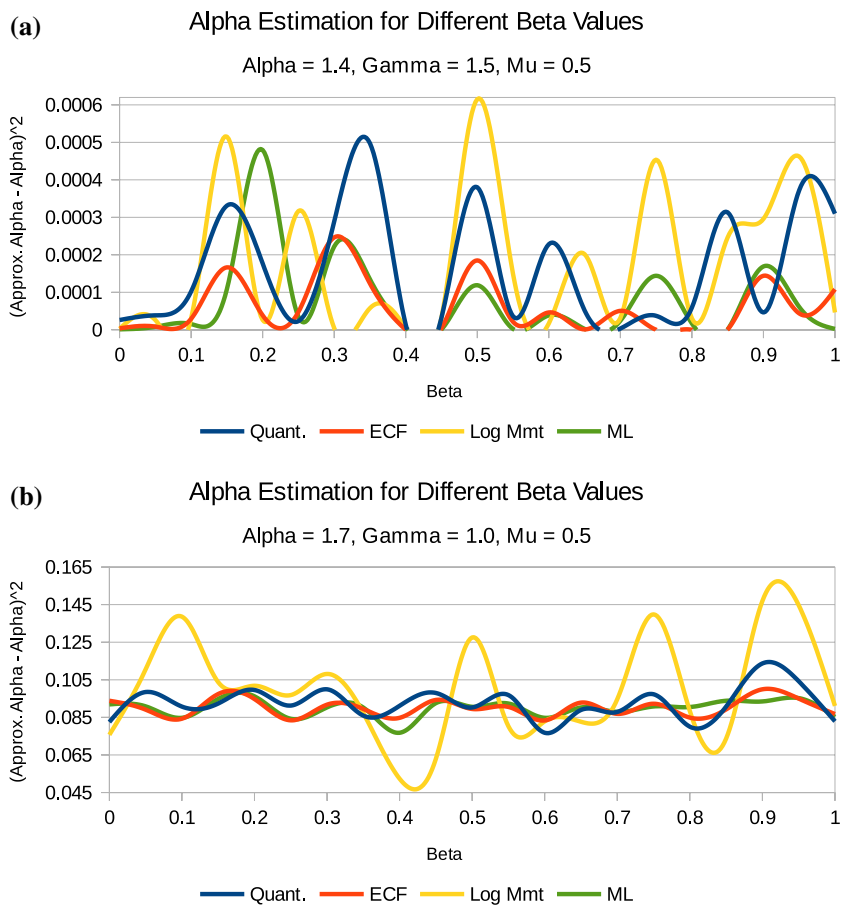

The graphs in Figure 4 illustrate convergence of the quantile, ECF and the MLE in estimating $\alpha = 1.4$ and $\alpha = 1.7$. We simulated 50,000 points and divided it into 100 sets starting with a 500-sized set and increasing it by 500 to 50,000. The logarithm moments method performed extremely poorly and incomparable to the above three methods. It is not included in Figure 4(a) and (b). The ECF is seen to perform better than the quantile and ML methods with a relatively better convergence rate. Similary Figure 5 shows the convergence rates for the quantile, ECF and ML estimation methods. The ECF still provides a better precision in both cases i.e. Figure 5(a) and (b).

**Figure 4. $\beta$ estimation for differing data set sizes and $\alpha$ values.**

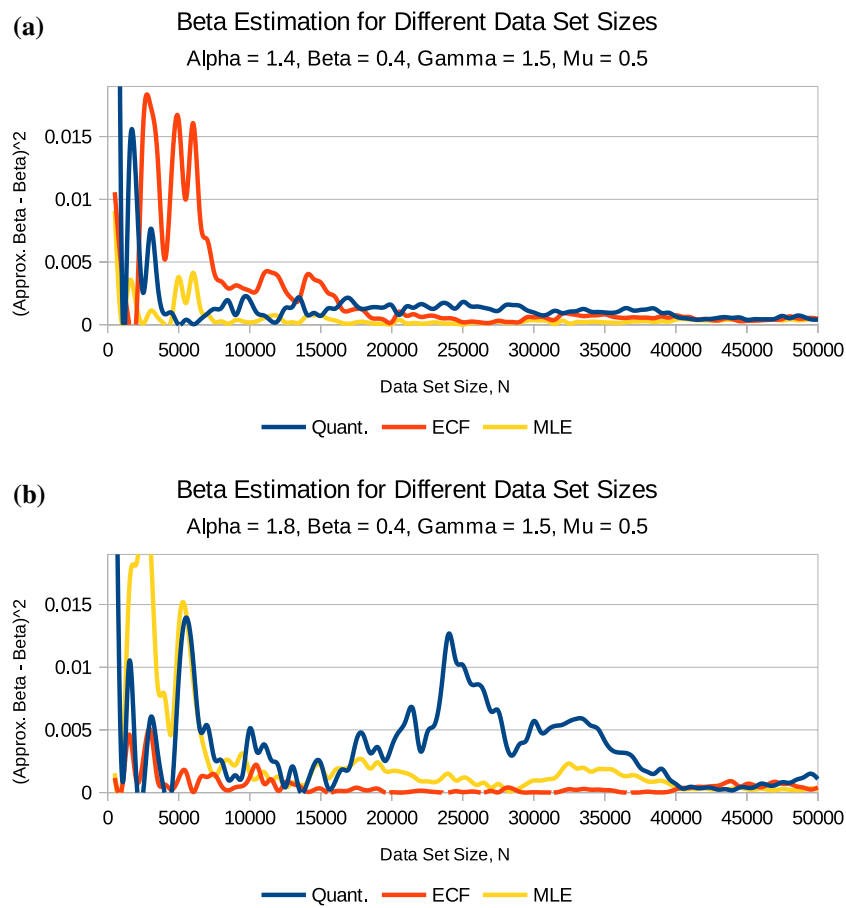

cogent ·· economics & finance

**Figure 5.** α estimation for differing data set sizes for β = 0.4 values.

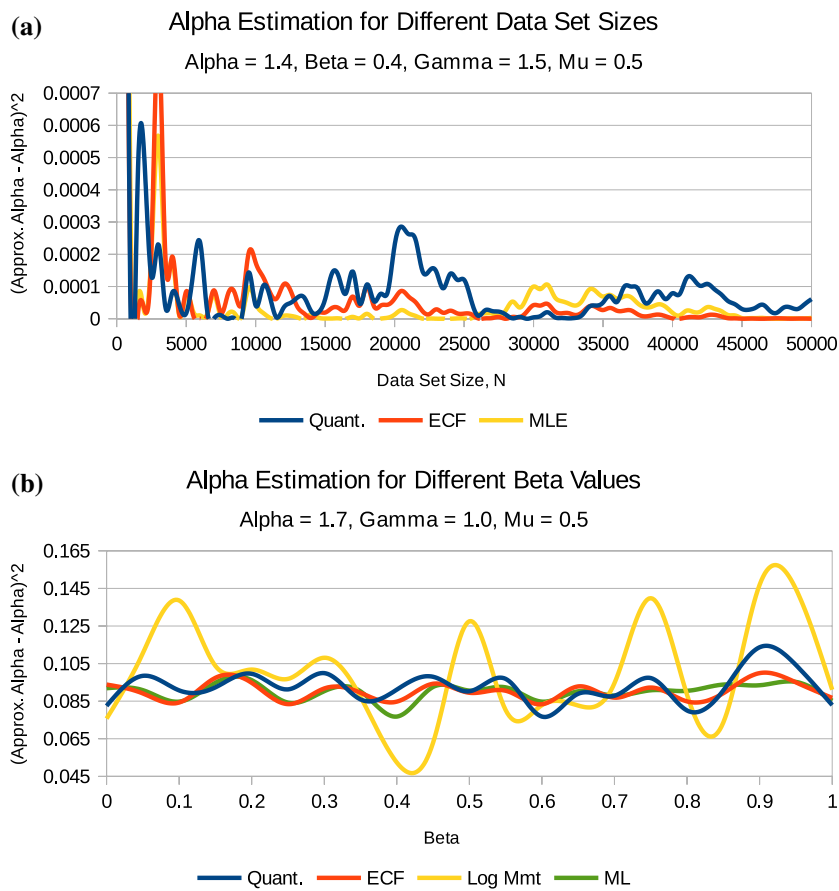

In summary the empirical characteristic function method outperforms all the three other methods discussed in this paper in the following way:

(1) It is robust and can consistently estimate a wide range of $\alpha$ and $\beta$ parameters.

(2) It provides a better precision compared to the quantile, logarithm moments and MLE methods for a wide range of $\alpha$ and $\beta$ parameters.

(3) It has a better convergence rate.

Therefore the quantile, logarithm moments or the ML methods can be used to provide initial parameters for the ECF method. Similarly, the latter can be used to provide initial parameters for better future estimators.

The following section is devoted to extracting stable parameters from log-returns commodity futures data using the ECF method.

## 5. Commodity data

The data sets used here are obtained from Quandl Financial and Economic Data website. The sets differ in sizes and include settled prices of Corn Futures Continuous Contract C#1 from 1959-07-01 to 2017-02-10; Crude Oil Futures Continuous Contract C#1 from 1983-03-30 to 2017-02-10; Gasoline Futures Continuous Contract C#1 from 2005-10-03 to 2017-02-10; Gold Futures Continuous Contract C#1 from 1974-12-31 to 2017-02-10; Natural Gas Futures Continuous Contract C#1 from 1990-04-03 to 2017-02-10; Platinum Futures Continuous Contract C#1 from 1969-01-02 to 2017-02-10; Silver Futures Continuous Contract C#1 from 1963-06-13 to 2017-02-10; Soybeans Futures Continuous Contract C#1 from 1959-07-01 to 2017-02-10; Wheat Futures Continuous Contract C#1 from 1959-07-01 to 2017-02-10. To avoid multi-distributional effects, we work with log-returns of the data sets.

### 5.1. The t-location-scale distribution

The *t*-location-scale distribution is most suited for modeling data distributions with heavier tails, more prone to outliers than the Gaussian distribution. The distribution uses the following parameters

| Parameter | Description | Support |
|-----------|-------------|---------|
| $\mu$ | Location parameter | $-\infty < \mu < \infty$ |
| $\nu$ | Scale parameter | $\nu > 0$ |
| $\alpha^*$ | Shape parameter | $\alpha^* > 0$ |

The probability density function (pdf) of the *t*-location-scale distribution is given by

$$h(x) = \frac{\Gamma\left(\frac{\alpha^*+1}{2}\right)}{\nu\sqrt{\alpha^*\pi}\Gamma\left(\frac{\alpha^*}{2}\right)}\left(\frac{\alpha^* + \left(\frac{x-\mu}{\nu}\right)^2}{\alpha^*}\right)^{-\left(\frac{\alpha^*+1}{2}\right)},$$

where $\Gamma(\cdot)$ denotes the gamma function. The mean of the *t*-location-scale distribution is given by $\mu$ and it is defined for $\alpha^* > 1$ and undefined otherwise. The variance is given by

$$\mathbb{V}ar = \nu^2\frac{\alpha^*}{\alpha^* - 2}.$$

The *t*-location-scale distribution approaches the Gaussian distribution as $\alpha^*$ approaches infinity and smaller values of $\alpha^*$ yield heavier tails. This distribution does not take skewness into consideration and its three parameters are usually estimated using the ML estimation method.

Using algorithms by Sheppard (2012) on our log-returns commodity futures data we obtained fittings in Figures 6–8.

According to the $\alpha^*$ values, the log-returns data exhibit some tails. To determine the nature of the details one would require to run some QQ plots but this can also be observed directly from the Figures 6–8.

cogent •• economics & finance

**Figure 6. Energy: The data exhibits high peaks and skinny tails.**

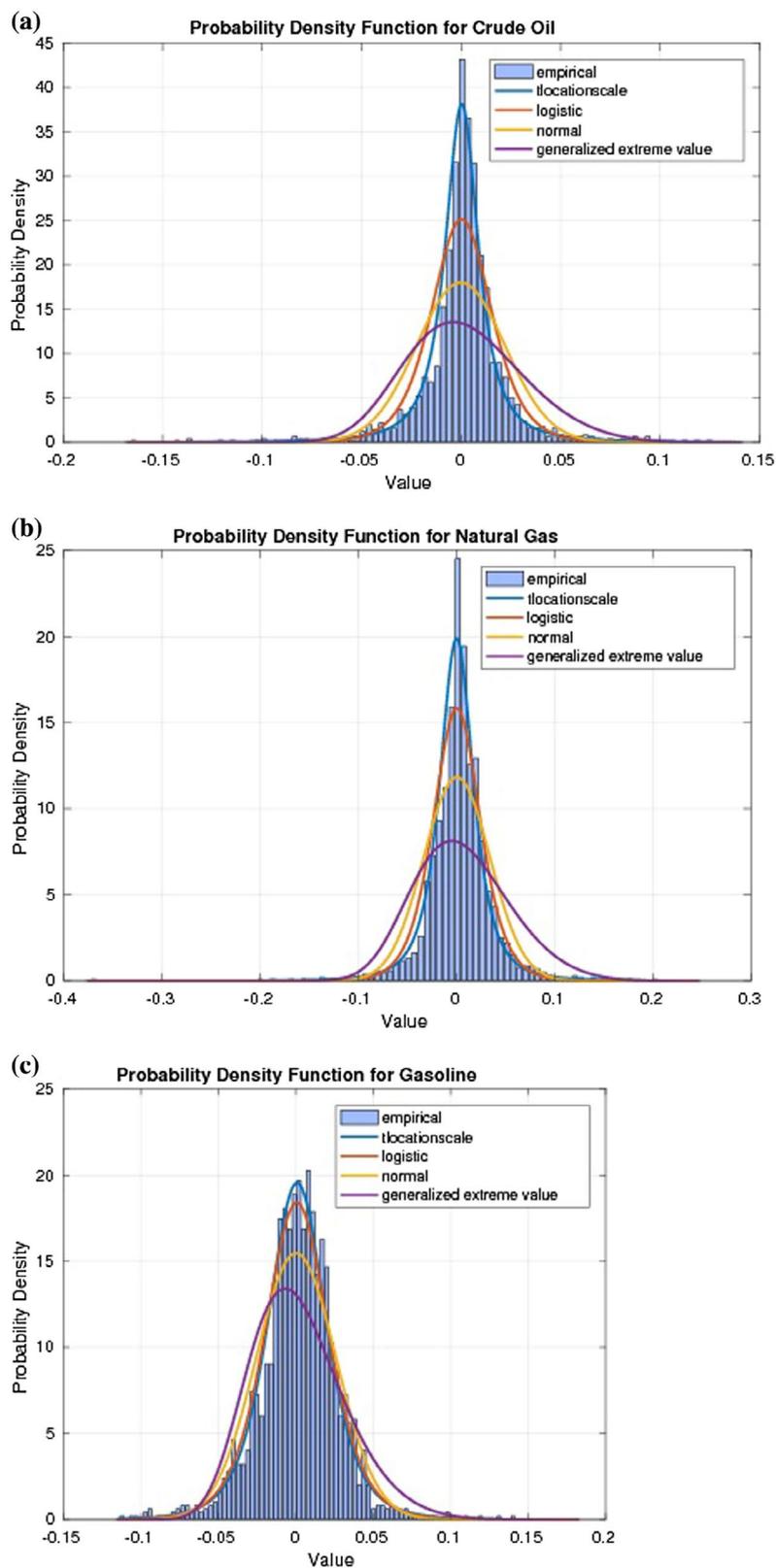

cogent •• economics & finance

**Figure 7. Grains: The data exhibits high peaks and skinny tails.**

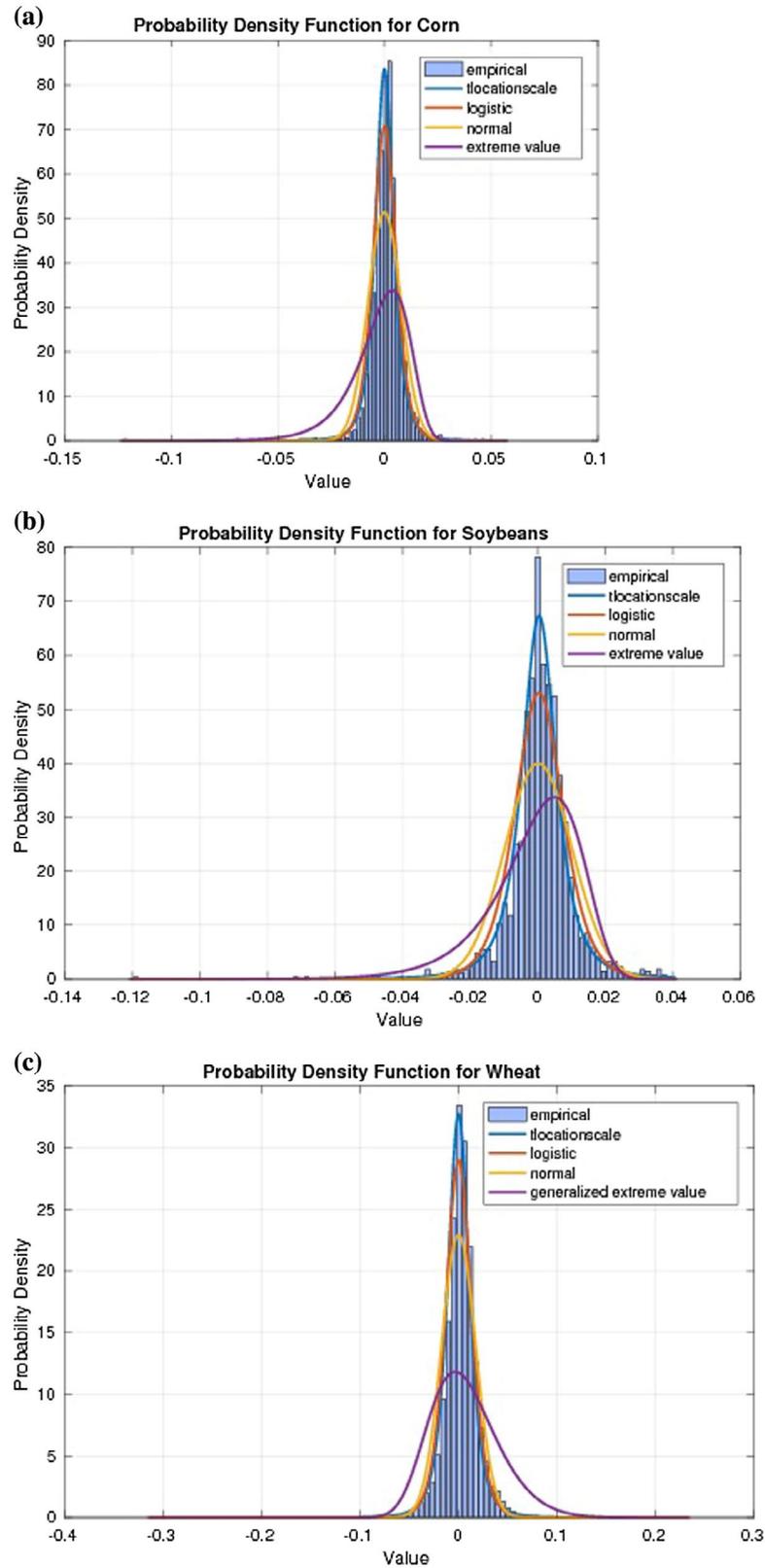

**Figure 8. Metals: The data exhibits high peaks and skinny tails.**

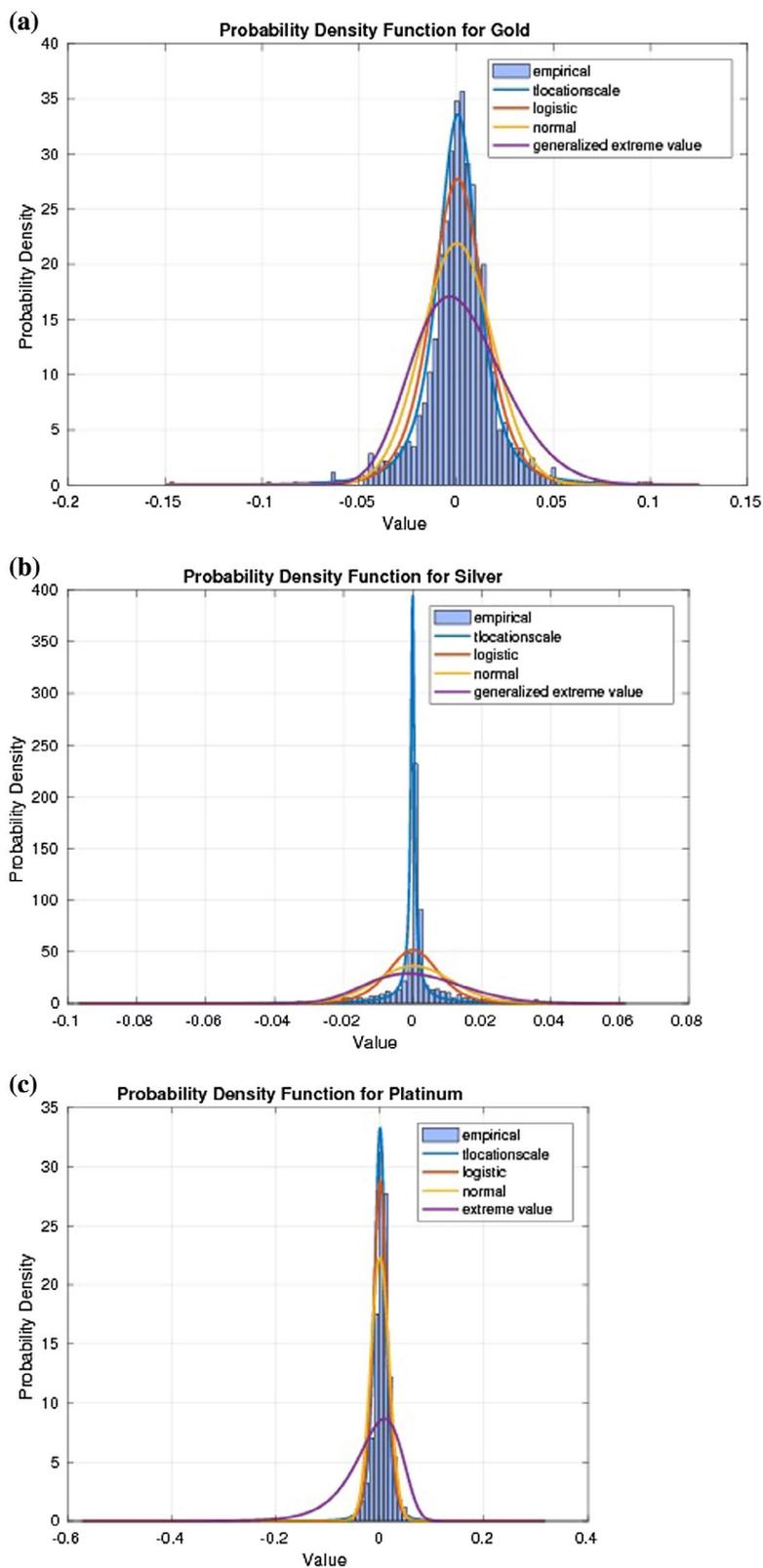

cogent •• economics & finance

| Table 2. *t*-location-scale distribution parameters extracted from the log-returns data | | | | |
|---|---|---|---|---|
| | | **t-location scale parameters** | | |
| | | **$\mu$** | **$\nu$** | **$\alpha^*$** |
| Energy | Gasoline | 0.000905848 | 0.0193048 | 4.45922 |
| | Natural gas | 0.000117668 | 0.0181944 | 2.50848 |
| | Crude oil | 0.000313708 | 0.00912801 | 1.75246 |
| Grains | Corn | 5.52294e − 05 | 0.00439924 | 3.03782 |
| | Soy beans | 0.000400308 | 0.00535733 | 2.43218 |
| | Wheat | 1.63112e − 05 | 0.0113397 | 3.41863 |
| Metals | Gold | 0.000905944 | 0.0108491 | 2.70553 |
| | Platinum | 0.00044219 | 0.0110761 | 3.13631 |
| | Silver | −1.72459e − 05 | 0.000682631 | 0.512196 |

It is important to mention that QQ plots do not straight away provide conclusive evidence about the nature of the tails. More tests would still need to be made. For instance under the *t*-location scale it is not obvious to observe any skewness in the data. We however, view this effect when we fit the data to stable distribution (see Table 2) as discussed in the following section.

### 5.2. Stable distribution fitting
On the other hand, by assuming stable distribution for our log-returns commodity futures data, we employed the ECF method and obtained the stable parameters in Table 3.

Log-returns of commodity futures are not only high peaked but they also have left and right skinny tails with extreme outliers as observed from the QQ-plots for energy commodities (i.e. Crude oil, Natural gas and Gasoline) in Figure 9, the grains commodities in Figure 10 and the precious metals in Figure 11.

Table 3 shows stable distribution parameters extracted from the log-returns data using the empirical characteristic function parameter estimation method. We notice that the data exhibit a bit of skewness which is not reflected in the *t*-location-scale distribution fitting.

| Table 3. Stable distribution parameters extracted from the log-returns data | | | | | |
|---|---|---|---|---|---|
| | | **Stable distribution parameters** | | | |
| | | **$\alpha$** | **$\beta$** | **$\nu$** | **$\mu$** |
| Energy | Gasoline | 1.7504 | −0.3806 | 0.0152 | −0.0005 |
| | Natural gas | 1.5329 | 0.0371 | 0.015 | 0.0005 |
| | Crude oil | 1.2322 | −0.1526 | 0.0075 | −0.0022 |
| Grains | Corn | 1.651 | 0.2117 | 0.0036 | 0.0004 |
| | Soy beans | 1.4665 | −0.0968 | 0.0043 | 0.0001 |
| | Wheat | 1.638 | 0.0929 | 0.0091 | 0.0003 |
| Metals | Gold | 1.5007 | −0.1324 | 0.0088 | 0.0001 |
| | Platinum | 1.5943 | −0.1339 | 0.0089 | −0.0001 |
| | Silver | 0.4461 | 0.0176 | 0.0011 | −0.0001 |

**Figure 9. Energy: In all, the left and right tails are skinny.**

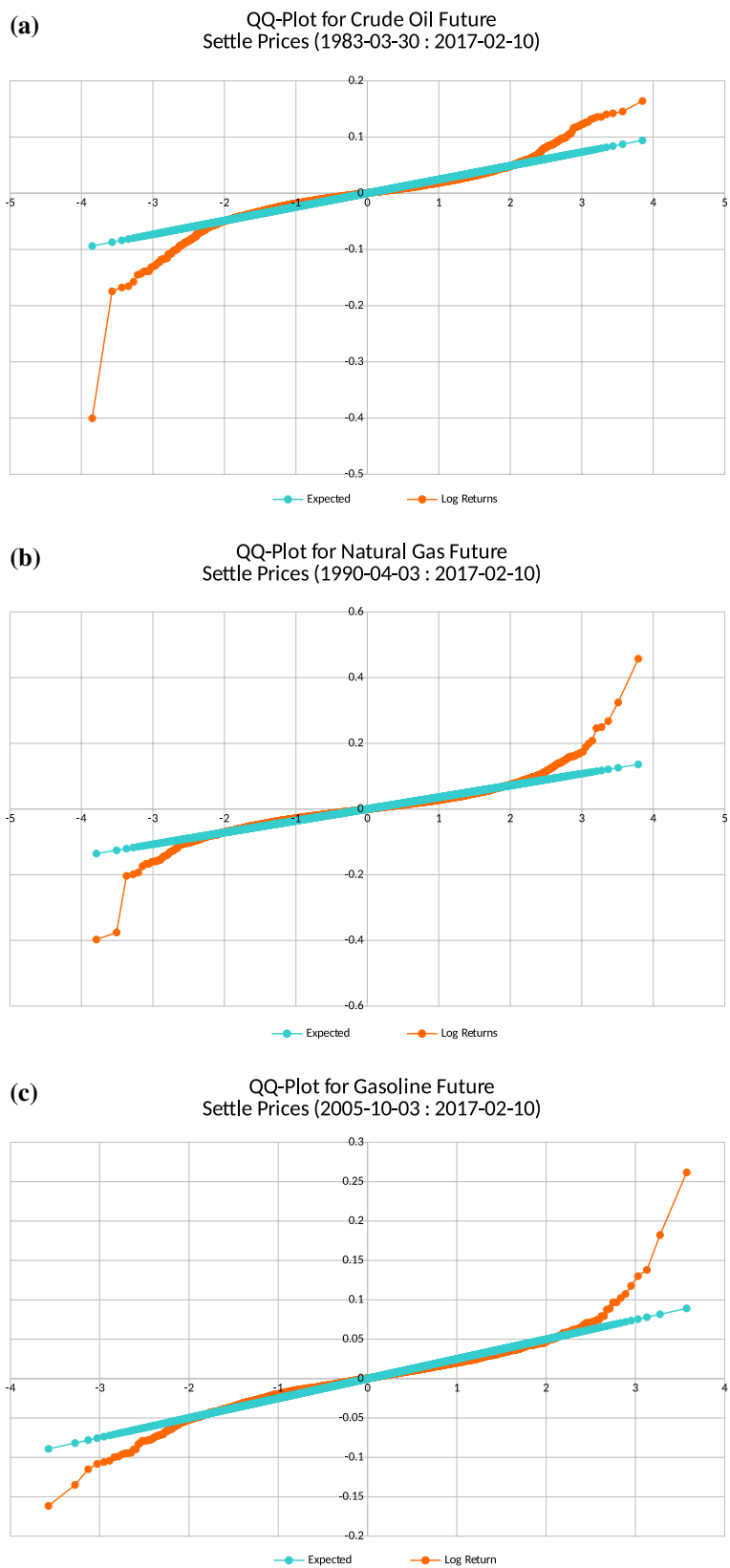

cogent •• economics & finance

**Figure 10. Grains: In all, the left and right tails are skinny.**

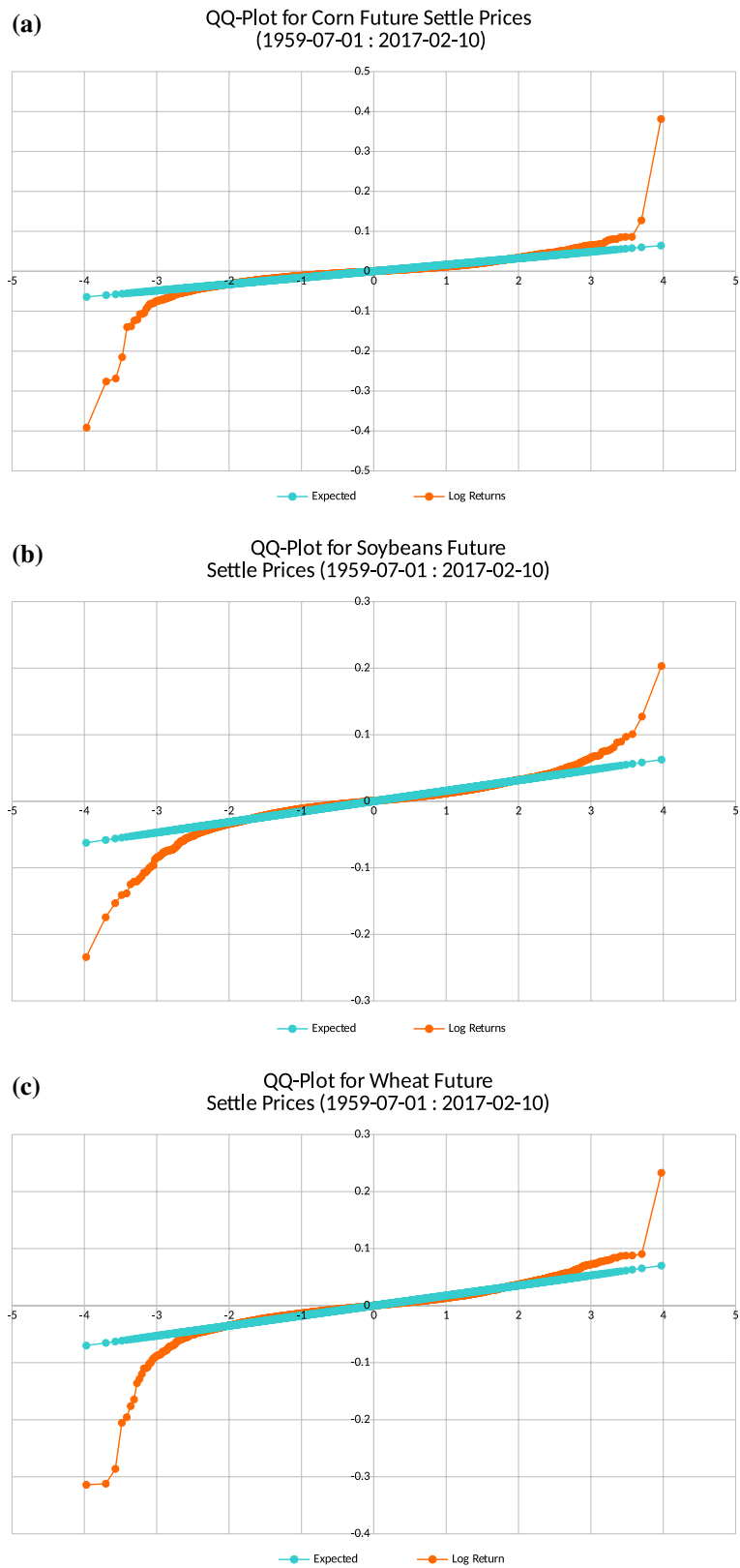

cogent ·· economics & finance

**Figure 11. Metals: In all, the left and right tails are skinny.**

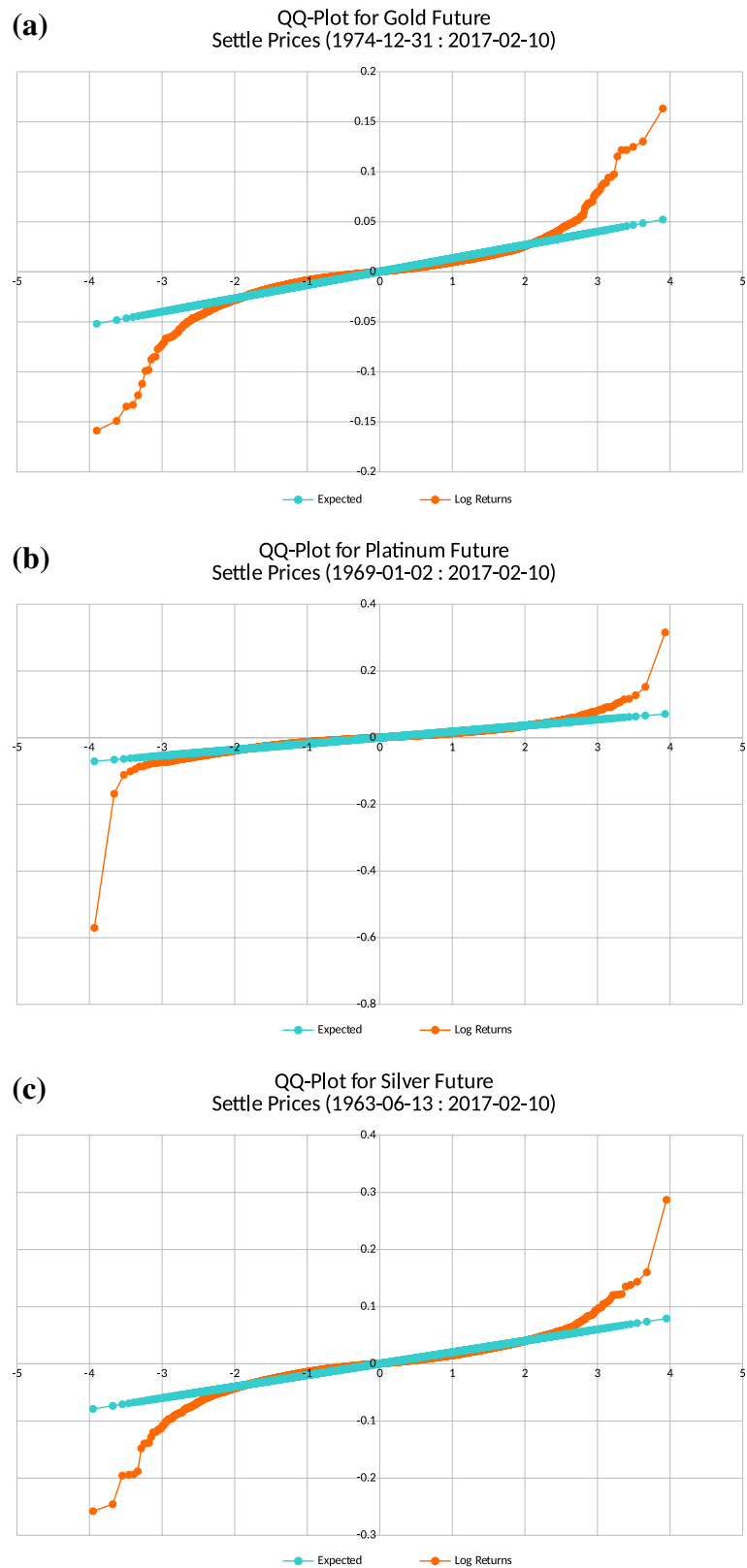

cogent •• economics & finance

## 6. Conclusion

First we showed that the ECF provides the best precision in estimating a wide range of $\alpha$ and $\beta$ parameters, it is robust and provides better convergence compared to the quantile, ML, and the logarithm moments. Secondly, we have illustrated that in general, the distribution of the commodity futures log-returns data is closest to a $t$-location-scale distribution due to its high peaks, skinny tails and extreme outliers. Moreover, by using the ECF estimation method we realize some minor skewness effects not captured in the $t$-location-scale fitting. We recommend the ECF as a suitable approach for estimating parameters of any skewed financial market data and could be used to obtain initial input parameters for future and better estimation techniques.

**Funding**
This work was supported by funds from the National Research Foundation of South Africa (NRF), the African Institute for Mathematical Sciences (AIMS) and the African Collaboration for Quantitative Finance and Risk Research (ACQuFRR) which is the research section of the African Institute of Financial Markets and Risk Management (AIFMRM), which delivers postgraduate education and training in financial markets, risk management and quantitative finance at the University of Cape Town in South Africa.

**Author details**
M. Kateregga[1]
E-mail: michaelk@aims.ac.za
S. Mataramvura[1]
E-mail: sure.mataramvura@uct.ac.za
ORCID ID: http://orcid.org/0000-0002-8073-2070
D. Taylor[1]
E-mail: david.taylor@uct.ac.za
[1] Actuarial Science, University of Cape Town, Rondebosch, Cape Town 7700, South Africa.

**Notes**
1. Note that characteristic functions always exist.
2. These are easily obtained from in-built functions in MATLAB

**Cover image**
Source: Original image from Parameter estimation for stable distributions with application to commodity futures log-returns by M. Kateregga, S. Mataramvura and D. Taylor.

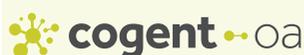

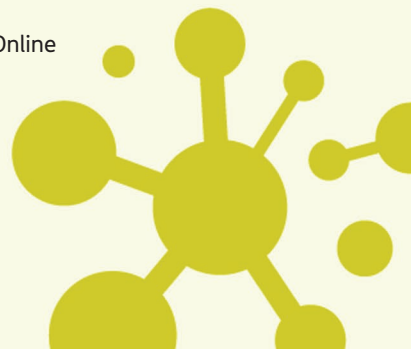